\documentclass[twocolumn,preprintnumbers,pra,hyperref]{revtex4}
\usepackage{amssymb}
\usepackage{amsfonts}
\usepackage{amsmath}
\usepackage{graphicx}

\begin{document}

\title{New approach for normalization and photon-number distributions of photon-added
(-subtracted) squeezed thermal states}
\author{Li-Yun Hu$^{1,2\dag }$ and Zhi-Ming Zhang$^{2\ddagger }$}
\affiliation{$^{1}${\small College of Physics \& Communication Electronics, Jiangxi
Normal University, Nanchang 330022, China}\\
$^{2}${\small Key Laboratory of Photonic Information Technology of Guangdong
Higher Education Institutes, }\\
{\small SIPSE \& LQIT, South China Normal University, Guangzhou 510006, China%
}\\
$\dag ${\small E-mail: hlyun2008@126.com; \ddag\ E-mail: zmzhang@scnu.edu.cn.%
}}

\begin{abstract}
{\small Using the thermal field dynamics theory to convert the thermal state
to a \textquotedblleft pure\textquotedblright\ state in doubled Fock space,
it is found that the average value of }$e^{fa^{\dag }a}$ {\small under
squeezed thermal state (STS) is just the generating function of Legendre
polynomials, a remarkable result. Based on this point, the normalization and
photon-number distributions of m-photon added (or subtracted) STS are
conviently obtained as the Legendre polynomials. This new concise method can
be expanded to the entangled case.}
\end{abstract}

\maketitle

\section{Introduction}

Nonclassicality of optical fields is helpful in understanding fundamentals
of quantum optics and have many applications in quantum information
processing \cite{1}. To generate and manipulate various nonclassical optical
fields, subtracting or adding photons from/to traditional quantum states or
Gaussian states is proposed \cite{2,2a,3,4,5,6,7,8}. For example, the photon
addition and subtraction have been successfully demonstrated experimentally
for probing quantum commutation rules by Parigi \textit{et al}. \cite{6}.
Recently, photon-added (-subtracted) Gaussian states have received more
attention from both experimentalists and theoreticians \cite%
{9,10,11,11a,12,13,14,15,16,17}, since these states exhibit an abundant of
nonclassical properties and may give access to a complete engineering of
quantum states and to fundamental quantum phenomena.

Theoretically, the normalization factors of such quantum states are
essential for studying their nonclassical properties. Very recent, Fan and
Jiang \cite{18} present a new concise approach for normalizing
m-photon-added (-subtracted) squeezed vacuum state (pure state) by
constructing generating function. However, most systems are not isolated,
but are immersed in a thermal reservoir, thus it is often the case that we
have no enough information to specify completely the state of a system. In
such a situation, the system only can be described by mixed states, such as
thermal states. In addition, the squeezed thermal states (STSs) can be
considered as the generalized Gaussian states.

In this paper, we shall extend this case to the mixed state, i.e., by using
the thermal field dynamics (TFD) theory to convert the thermal state to a
\textquotedblleft pure\textquotedblright\ state in doubled Fock space, we
present a new concise method for normalizing photon-added (-subtracted)
squeezed thermal states (PASTSs, PSSTSs) and for deriving their
photon-number distributions (PNDs) which have been a major topic of studies
on quantum optics and quantum statistics. It is found that the normalization
factors and PNDs are related to the Legendre polynomials in a compact form%
{\small .}

Our paper is arranged as follows. In section 2, based on Takahashi-Umezawa
TFD, we convert the thermal state $\rho _{th}$ to a pure state in doubled
Fock space by the partial trace, $\rho _{th}=\mathtt{t\tilde{r}}\left[
\left\vert 0(\beta )\right\rangle \left\langle 0(\beta )\right\vert \right] $
(see Eq.(\ref{1.4}) below). In section 3, we introduce the STS $%
S_{1}\left\vert 0(\beta )\right\rangle $ with $S_{1}$ being the single-mode
squeezing operator for the real mode. It is shown that the average value of $%
e^{fa^{\dag }a}$ under STS is just the generating function of Legendre
polynomials, a remarkable result. Based on this point, in sections 4 and 5,
the normalization factors and PNDs of m-photon added (or subtracted) STS are
obtained as the Legendre polynomials, respectively. The last section is
devoted to drawing a conclusion.

\section{Representation of thermal state in doubled Fock space}

We begin with briefly reviewing the properties of thermal state. For a
single field mode with frequency $\omega $ in a thermal equilibrium state
corresponding to absolute temperature $T$, the density operator is
\begin{equation}
\rho _{th}=\sum_{n=0}^{\infty }\frac{n_{c}^{n}}{\left( n_{c}+1\right) ^{n+1}}%
\left\vert n\right\rangle \left\langle n\right\vert ,  \label{1.1}
\end{equation}%
where $n_{c}=[\exp (\hbar \omega /(kT))-1]^{-1}$ being the average photon
number of the thermal state $\rho _{th}$ and $k$ being Bltzmann's constant.
Note $\left\vert n\right\rangle =a^{\dagger n}/\sqrt{n!}\left\vert
0\right\rangle $ and the normally ordering form of vacuum projector $%
\left\vert 0\right\rangle \left\langle 0\right\vert =\colon \exp
(-a^{\dagger }a)\colon $(the symbol $\colon \colon $ denotes the normal
ordering), one can put Eq.(\ref{1.1}) into the following form%
\begin{equation}
\rho _{th}=\colon \frac{1}{n_{c}+1}e^{-\frac{1}{n_{c}+1}a^{\dag }a}\colon =%
\frac{1}{n_{c}+1}e^{a^{\dag }a\ln \frac{n_{c}}{n_{c}+1}},  \label{1.2}
\end{equation}%
where in the last step, the operator identity $\exp \left( \lambda a^{\dag
}a\right) =\colon \exp \left[ \left( e^{\lambda }-1\right) a^{\dag }a\right]
\colon $is used.

Recalling the thermal field dynamics (TFD) introduced by Takahashi and
Umezawa \cite{19,20,21}, its elemental spirit is to convert the calculations
of ensemble averages for a mixed state $\rho $, $\left\langle A\right\rangle
=\mathtt{tr}\left( A\rho \right) /\mathtt{tr}\left( \rho \right) ,$ to
equivalent expectation values with a pure state $\left\vert 0(\beta
)\right\rangle $, i.e.,%
\begin{equation}
\left\langle A\right\rangle =\left\langle 0(\beta )\right\vert A\left\vert
0(\beta )\right\rangle ,  \label{1.3}
\end{equation}%
where $\beta =1/kT$, $k$ is the Boltzmann constant. Thus, for the density
operator $\rho _{th}$, by using the partial trace method \cite{22}, i.e., $%
\rho _{th}=\widetilde{\mathtt{tr}}\left[ \left\vert 0(\beta )\right\rangle
\left\langle 0(\beta )\right\vert \right] $ where $\widetilde{\mathtt{tr}}$
denotes the trace operation over the environment freedom (denoted as
operator $\tilde{a}^{\dag }$), one can obtain the explicit expression of $%
\left\vert 0(\beta )\right\rangle $ in doubled Fock space,
\begin{equation}
\left\vert 0(\beta )\right\rangle =\text{sech}\theta \exp \left[ a^{\dagger }%
\tilde{a}^{\dagger }\tanh \theta \right] \left\vert 0\tilde{0}\right\rangle
=S\left( \theta \right) \left\vert 0\tilde{0}\right\rangle ,  \label{1.4}
\end{equation}%
where $\left\vert 0\tilde{0}\right\rangle $ is annihilated by $\tilde{a}$
and $a$, $[\tilde{a},\tilde{a}^{\dag }]=1$, and $S\left( \theta \right) $ is
the thermal operator, $S\left( \theta \right) \equiv \exp \left[ \theta
\left( a^{\dagger }\tilde{a}^{\dagger }-a\tilde{a}\right) \right] \ $with a
similar form to the a two-mode squeezing operator except for the tilde mode,
and $\theta $ is a parameter related to the temperature by $\tanh \theta
=\exp \left( -\frac{\hbar \omega }{2kT}\right) $. $\left\vert 0(\beta
)\right\rangle $ is named thermal vacuum state.

Let $\mathtt{Tr}$ denote the trace operation over both the system freedom
(expressed by $\mathtt{tr}$) and the environment freedom by $\widetilde {%
\mathtt{tr}}$, i.e., $\mathtt{Tr}=\mathtt{tr}\widetilde{\mathtt{tr}}$, then
we have%
\begin{align}
\mathtt{tr}\left( A\rho_{th}\right) & =\mathtt{Tr}\left[ A\left\vert
0(\beta)\right\rangle \left\langle 0(\beta)\right\vert \right]  \notag \\
& =\mathtt{tr}\left[ A\widetilde{\mathtt{tr}}\left\vert 0(\beta
)\right\rangle \left\langle 0(\beta)\right\vert \right] ,  \label{1.5}
\end{align}
and the average photon number of the thermal state $\rho_{th}$ is
\begin{equation}
n_{c}=\mathtt{Tr}\left[ a^{\dagger}a\left\vert 0(\beta)\right\rangle
\left\langle 0(\beta)\right\vert \right] =\sinh^{2}\theta.
\end{equation}
Here we should emphasize that $\widetilde{\mathtt{tr}}\left\vert
0(\beta)\right\rangle \left\langle 0(\beta)\right\vert \neq\left\langle
0(\beta)\right\vert \left. 0(\beta)\right\rangle ,$ since $\left\vert
0(\beta)\right\rangle $ involves both real mode $a$ and fictitious mode $%
\tilde{a}$. From Eqs.(\ref{1.3}) and (\ref{1.4}) one can see that the
worthwhile convenience in Eq.(\ref{1.4}) is at the expense of introducing a
fictitious field (or called a tilde-conjugate field) in the extended Hilbert
space, i.e., the original optical field state $\left\vert n\right\rangle $\
in the Hilbert space $\mathcal{H}$\ is accompanied by a tilde state $%
\left\vert \tilde{n}\right\rangle $\ in $\mathcal{\tilde{H}}$. A similar
rule holds for operators: every Bose annihilation operator $a$\ acting on $%
\mathcal{H}$\ has an image $\tilde{a}$\ acting on $\mathcal{\tilde{H}}$.
These operators in $\mathcal{H}$ are commutative with those in $\mathcal{%
\tilde{H}}$.

\section{Suqeezed thermal vacuum state}

To realize our purpose, we first introduce the squeezed thermal vacuum
state, defined as $S_{1}\left( r\right) \left\vert 0(\beta )\right\rangle $,
where $S_{1}\left( r\right) =\exp [r(a^{2}-a^{\dagger 2})/2]$ is the
single-mode squeezing operator for the real mode with $r$ being squeezing
parameter. Note that Eq.(\ref{1.4}) and the Baker-Hausdorf lemma
\begin{equation}
S_{1}\left( r\right) a^{\dag }S_{1}^{\dag }\left( r\right) =a^{\dag }\cosh
r+a\sinh r,  \label{1.6}
\end{equation}%
then we get%
\begin{align}
S_{1}\left( r\right) \left\vert 0(\beta )\right\rangle & =S_{1}\left(
r\right) \text{sech}\theta \exp \left[ a^{\dagger }\tilde{a}^{\dagger }\tanh
\theta \right] \left\vert 0\tilde{0}\right\rangle   \notag \\
& =\text{sech}\theta \text{sech}^{1/2}r\exp \left[ \left( a^{\dag }\cosh
r+a\sinh r\right) \tilde{a}^{\dagger }\tanh \theta \right]   \notag \\
& \times \exp \left[ -\frac{a^{\dagger 2}}{2}\tanh r\right] \left\vert 0%
\tilde{0}\right\rangle ,  \label{1.7}
\end{align}%
where we have used $S_{1}\left( \lambda \right) \left\vert 0\right\rangle =$%
sech$^{1/2}\lambda \exp [-a^{\dagger 2}/2\tanh \lambda ]\left\vert
0\right\rangle .$ Further, note $e^{\tau a\tilde{a}^{\dagger }}a^{\dagger
}e^{-\tau a\tilde{a}^{\dagger }}=a^{\dagger }+\tau \tilde{a}^{\dagger },$
and for operators $A,B$ satisfying the conditions $\left[ A,[A,B]\right] =%
\left[ B,[A,B]\right] =0,$ we have $e^{A+B}=e^{A}e^{B}e^{-[A,B]/2},$ thus
Eq.(\ref{1.7}) can be put into the following form%
\begin{align}
S_{1}\left( r\right) \left\vert 0(\beta )\right\rangle & =\text{sech}\theta
\text{sech}^{1/2}r\exp \left\{ \frac{\tanh \theta }{\cosh r}a^{\dag }\tilde{a%
}^{\dagger }\right.   \notag \\
& +\left. \frac{\tanh r}{2}\left( \tilde{a}^{\dagger 2}\tanh ^{2}\theta
-a^{\dagger 2}\right) \right\} \left\vert 0\tilde{0}\right\rangle .
\label{1.8}
\end{align}

Next, we shall use Eq.(\ref{1.8}) to derive the average of operator $%
e^{fa^{\dagger }a}$ under the suqeezed thermal vacuum state $S_{1}\left(
r\right) \left\vert 0(\beta )\right\rangle $, which is a bridge for our
whole calculations. Notice $e^{f/2a^{\dagger }a}a^{\dagger
}e^{-f/2a^{\dagger }a}=a^{\dagger }e^{f/2}$ and $e^{-f/2a^{\dagger
}a}ae^{f/2a^{\dagger }a}=ae^{f/2}$, so we have%
\begin{align}
e^{f/2a^{\dagger }a}S_{1}\left( r\right) \left\vert 0(\beta )\right\rangle &
=\text{sech}\theta \text{sech}^{1/2}r\exp \left\{ \frac{\tanh \theta }{\cosh
r}a^{\dag }\tilde{a}^{\dagger }e^{f/2}\right.   \notag \\
& +\left. \frac{\tanh r}{2}\left( \tilde{a}^{\dagger 2}\tanh ^{2}\theta
-a^{\dagger 2}e^{f}\right) \right\} \left\vert 0\tilde{0}\right\rangle ,
\label{1.9}
\end{align}%
which leads to%
\begin{align}
\left\langle e^{fa^{\dagger }a}\right\rangle & \equiv \left\langle 0(\beta
)\right\vert S_{1}^{\dag }\left( r\right) e^{fa^{\dagger }a}S_{1}\left(
r\right) \left\vert 0(\beta )\right\rangle   \notag \\
& =\text{sech}^{2}\theta \text{sech}r\left\langle 0\tilde{0}\right\vert \exp
\left\{ \frac{e^{f/2}\tanh \theta }{\cosh r}a\tilde{a}\right.   \notag \\
& +\left. \frac{\tanh r}{2}\left( \tilde{a}^{2}\tanh ^{2}\theta
-a^{2}e^{f}\right) \right\}   \notag \\
& \times \exp \left\{ \frac{\tanh \theta }{\cosh r}a^{\dag }\tilde{a}%
^{\dagger }e^{f/2}\right.   \notag \\
& +\left. \frac{\tanh r}{2}\left( \tilde{a}^{\dagger 2}\tanh ^{2}\theta
-a^{\dagger 2}e^{f}\right) \right\} \left\vert 0\tilde{0}\right\rangle
\notag \\
& =\left[ Ce^{2f}-2Be^{f}+A\right] ^{-1/2},  \label{1.10}
\end{align}%
where we have set%
\begin{align}
A& =\cosh ^{4}\theta +\cosh 2\theta \sinh ^{2}r  \notag \\
& =n_{c}^{2}+\left( 2n_{c}+1\right) \cosh ^{2}r,  \notag \\
B& =\sinh ^{2}\theta \cosh ^{2}\allowbreak \theta =n_{c}\left(
n_{c}+1\right) ,  \notag \\
C& =\cosh ^{4}\theta -\cosh 2\theta \cosh ^{2}r  \notag \\
& =n_{c}^{2}-\left( 2n_{c}+1\right) \sinh ^{2}r,  \label{1.10b}
\end{align}%
and we have used the completness relation of coherent state $\int d^{2}zd^{2}%
\tilde{z}\left\vert z\tilde{z}\right\rangle \left\langle z\tilde{z}%
\right\vert /\pi ^{2}=1,$ here $\left\vert z\right\rangle $ and $\left\vert
\tilde{z}\right\rangle $ is the coherent state in real and fictitious modes,
respectively, and the formula \cite{23}
\begin{align}
& \int \frac{d^{2}z}{\pi }\exp \left( \zeta \left\vert z\right\vert ^{2}+\xi
z+\eta z^{\ast }+fz^{2}+gz^{\ast 2}\right)   \notag \\
& =\frac{1}{\sqrt{\zeta ^{2}-4fg}}\exp \left[ \frac{-\zeta \xi \eta +\xi
^{2}g+\eta ^{2}f}{\zeta ^{2}-4fg}\right] ,  \label{1.11}
\end{align}%
whose convergent condition is Re$\left( \zeta \pm f\pm g\right) <0,\ $Re($%
\zeta ^{2}-4fg)/(\zeta \pm f\pm g)<0.$ Eq.(\ref{1.10}) is very important for
later calculation of photon-number distribution (PND) and normalization of
photon-added (-subtracted) squeezed thermal states (PASTS, PSSTS).

It is interesting to notice that the standard generating function of
Legendre polynomials \cite{25} is given by

\begin{equation}
\frac{1}{\sqrt{1-2xt+t^{2}}}=\sum_{m=0}^{\infty}P_{m}\left( x\right) t^{m},
\label{1.17}
\end{equation}
thus comparing Eq.(\ref{1.10}) with Eq.(\ref{1.17}) we find%
\begin{align}
\left\langle e^{fa^{\dagger}a}\right\rangle & =A^{-1/2}\left[ \frac{C}{A}%
e^{2f}-2\frac{B}{A}e^{f}+1\right] ^{-1/2}  \notag \\
& =A^{-1/2}\sum_{m=0}^{\infty}P_{m}\left( B/\sqrt{AC}\right) \left( \sqrt{C/A%
}e^{f}\right) ^{m},  \label{1.17a}
\end{align}
which indicates that the average value of $e^{fa^{\dag}a}$ under squeezed
thermal state (STS) is just the generating function of Legendre polynomials,
a remarkable result. Next, we shall examine the normalizations and PNDs of
PASTS and PSSTS by using Eqs.(\ref{1.10}) and (\ref{1.17a}).

\section{Normalization and PND of PASTS}

The $m$-photon-added scheme, denoted by the mapping $\rho\rightarrow a^{\dag
m}\rho a^{m},$ was first proposed by Agarwal and Tara \cite{2}. Here, we
introduce the PASTS. Theoretically, the PASTS can be obtained by repeatedly
operating the photon creation operator $a^{\dagger}$ on a STS, so its
density operator is given by%
\begin{equation}
\rho_{ad}=C_{a,m}^{-1}a^{\dag m}S_{1}\rho_{th}S_{1}^{\dagger}a^{m},
\label{1.12}
\end{equation}
where $m$ is the added photon number (a non-negative integer), $C_{a,m}^{-1}$
is the normalization constant to be determined.

\subsection{Normalization of PASTS}

To fully describe a quantum state, its normalization is usually necessary.
Next, we shall employ the fact (\ref{1.5}) and (\ref{1.10}), (\ref{1.17a})
to realize our aim. According to the normalization condition \texttt{tr}$%
\rho _{ad}=1$ and the TFD, we have%
\begin{align}
C_{a,m}& =\mathtt{tr}\left[ a^{\dag m}S_{1}\rho _{th}S_{1}^{\dagger }a^{m}%
\right]   \notag \\
& =\mathtt{Tr}\left[ a^{\dag m}S_{1}\left\vert 0(\beta )\right\rangle
\left\langle 0(\beta )\right\vert S_{1}^{\dagger }a^{m}\right]   \notag \\
& =\left\langle 0(\beta )\right\vert S_{1}^{\dagger }a^{m}a^{\dag
m}S_{1}\left\vert 0(\beta )\right\rangle ,  \label{1.13}
\end{align}%
which implies that the calculation of normaliation factor $C_{a,m}$ is
converted to a matrix element after introducing the thermal vacuum state $%
\left\vert 0(\beta )\right\rangle $.

Note the operator identity \cite{24} $e^{\tau a^{\dag}a}=e^{-\tau}\vdots \exp%
[(1-e^{-\tau})a^{\dag}a]\vdots$, we see%
\begin{equation}
\sum_{m=0}^{\infty}\frac{\tau^{m}}{m!}a^{m}a^{\dag m}=\vdots e^{\tau a^{\dag
}a}\vdots=\left( \frac{1}{1-\tau}\right) ^{a^{\dag}a+1},  \label{1.14}
\end{equation}
where the symbol $\vdots$ $\vdots$ denotes antinormally ordering. Thus using
Eqs.(\ref{1.10}), (\ref{1.13}) and (\ref{1.14}), ($e^{f}\rightarrow\frac {1}{%
1-\tau}$) we have%
\begin{align}
\sum_{m=0}^{\infty}\frac{\tau^{m}}{m!}C_{a,m} & =\frac{1}{1-\tau }%
\left\langle 0(\beta)\right\vert S_{1}^{\dagger}e^{a^{\dag}a\ln\frac {1}{%
1-\tau}}S_{1}\left\vert 0(\beta)\right\rangle  \notag \\
& =\left[ A\tau^{2}-2D\tau+1\right] ^{-1/2},  \label{1.15}
\end{align}
where we have set%
\begin{align}
D & =\cosh^{2}\theta\cosh2r-\sinh^{2}r  \notag \\
& =n_{c}\cosh2r+\cosh^{2}r.  \label{1.16}
\end{align}

Comparing Eq.(\ref{1.15}) with Eq.(\ref{1.17}), and taking $\tau^{\prime
}\rightarrow\sqrt{A}\tau$, we obtain%
\begin{align}
\sum_{m=0}^{\infty}\tau^{\prime m}\frac{C_{a,m}}{m!A^{m/2}} & =\left[
\tau^{\prime2}-2D/\sqrt{A}\tau^{\prime}+1\right] ^{-1/2}  \notag \\
& =\sum_{m=0}^{\infty}P_{m}\left( D/\sqrt{A}\right) \tau^{\prime m},
\label{1.18}
\end{align}
thus the normalization constant of PASTS is given by
\begin{equation}
C_{a,m}=m!A^{m/2}P_{m}\left( D/\sqrt{A}\right) ,  \label{1.19}
\end{equation}
which is identical with the result in Ref.\cite{26}. It is noted that, for
the case of no-photon-addition with $m=0$, $C_{a,0}=1$ as expected. Under
the case of $m$-photon-addition thermal state (no squeezing) with $%
D=\allowbreak n_{c}+1$, $A=\allowbreak\left( n_{c}+1\right) ^{2},$ and $%
P_{m}\left( 1\right) =1$, then $C_{a,m}=m!\left( n_{c}+1\right) ^{m}.$ The
same result as Eq.(32) found in Ref.\cite{27}. In addition, when $r=0$
corresponding to photon-added thermal state, Eq.(\ref{1.19}) just reduces to
$C_{a,m}=m!\cosh^{2m}\theta$ \cite{27}.

\subsection{PND of PASTS}

The photon-number distribution (PND) is a key characteristic of every
optical field. The PND, i.e., the probability of finding $n$ photons in a
quantum state described by the density operator $\rho$, is $\mathcal{P}(n)=%
\mathtt{tr}\left[ \left\vert n\right\rangle \left\langle n\right\vert \rho%
\right] $. In a similar spirit of deriving Eq.(\ref{1.19}), noting $%
a^{m}\left\vert n\right\rangle =\sqrt{n!/(n-m)!}\left\vert n-m\right\rangle $
and $\left\vert n\right\rangle =a^{\dagger n}/\sqrt{n!}\left\vert
0\right\rangle ,$ the PND of the PASTS can be calculated as
\begin{align}
\mathcal{P}_{a}(n) & =C_{a,m}^{-1}\mathtt{tr}\left[ \left\vert
n\right\rangle \left\langle n\right\vert a^{\dag
m}S_{1}\rho_{th}S_{1}^{\dagger}a^{m}\right]  \notag \\
& =C_{a,m}^{-1}\mathtt{Tr}\left[ \left\vert n\right\rangle \left\langle
n\right\vert a^{\dag m}S_{1}\left\vert 0(\beta)\right\rangle \left\langle
0(\beta)\right\vert S_{1}^{\dagger}a^{m}\right]  \notag \\
& =\frac{n!C_{a,m}^{-1}}{l!}\left\langle 0(\beta)\right\vert S_{1}^{\dagger
}\left\vert l\right\rangle \left\langle l\right\vert S_{1}\left\vert
0(\beta)\right\rangle  \notag \\
& =\frac{n!C_{a,m}^{-1}}{\left( l!\right) ^{2}}\left\langle 0(\beta
)\right\vert S_{1}^{\dagger}a^{\dagger l}\left\vert 0\right\rangle
\left\langle 0\right\vert a^{l}S_{1}\left\vert 0(\beta)\right\rangle
\label{1.20}
\end{align}
which leads to
\begin{align}
& \sum_{l=0}^{\infty}\tau^{l}\frac{l!}{n!}C_{a,m}\mathcal{P}_{a}(n)  \notag
\\
& =\sum_{l=0}^{\infty}\frac{\tau^{l}}{l!}\left\langle 0(\beta)\right\vert
S_{1}^{\dagger}\colon\left( a^{\dagger}a\right) ^{l}e^{-a^{\dagger}a}\colon
S_{1}\left\vert 0(\beta)\right\rangle  \notag \\
& =\left\langle 0(\beta)\right\vert S_{1}^{\dagger}e^{a^{\dagger}a\ln\tau
}S_{1}\left\vert 0(\beta)\right\rangle ,  \label{1.21}
\end{align}
where $l=n-m$ and the vacuum projector operator $\left\vert 0\right\rangle
\left\langle 0\right\vert =\colon e^{-a^{\dagger}a}\colon$ and the operator
identity $e^{\lambda a^{\dagger}a}=\colon\exp[(e^{\lambda}-1)a^{\dagger }a]%
\colon$ are used.

Using Eq.(\ref{1.10}) again ($e^{f}\rightarrow\tau$) and comparing Eq.(\ref%
{1.21}) with Eq. (\ref{1.17}) we see%
\begin{align}
& \sum_{l=0}^{\infty}\tau^{l}\frac{l!}{n!}C_{a,m}\mathcal{P}_{a}(n)  \notag
\\
& =A^{-1/2}\left[ \frac{C}{A}\tau^{2}-2\frac{B}{A}\tau+1\right] ^{-1/2}
\notag \\
& =A^{-1/2}\sum_{l=0}^{\infty}P_{l}\left( B/\sqrt{AC}\right) \left( \sqrt{C/A%
}\tau\right) ^{l},  \label{1.22}
\end{align}
which leads to the PND of PASTS%
\begin{equation}
\mathcal{P}_{a}(n)=\frac{n!C_{a,m}^{-1}\left( C/A\right) ^{\left( n-m\right)
/2}}{\left( n-m\right) !\sqrt{A}}P_{n-m}\left( B/\sqrt {AC}\right) ,
\label{1.24}
\end{equation}
a Legendre polynomial with a condition $n\geqslant m$ which implies that the
photon-number ($n$) involved in PASTS is always no-less than the
photon-number ($m$) operated on the STS, and there is no photon distribution
when $n<m$)\textbf{. }It is obvious that when $m=0$ corresponding to the
STS, then the PND of STS is also a Legendre distribution \cite{28}.

\section{Normalization and PND of PSSTS}

Next, we turn our attention to discussing the photon-subtracted squeezed
thermal state (PSSTS), defined as%
\begin{equation}
\rho_{sb}=C_{s,m}^{-1}a^{m}S_{1}\rho_{th}S_{1}^{\dagger}a^{\dag m},
\label{1.25}
\end{equation}
where $m$ is the subtracted photon number (a non-negative integer), and $%
C_{s,m}$ is a normalized constant.

In a similar way to deriving Eq.(\ref{1.19}), we have%
\begin{equation}
C_{s,m}=\left\langle 0(\beta)\right\vert S_{1}^{\dagger}a^{\dag
m}a^{m}S_{1}\left\vert 0(\beta)\right\rangle ,  \label{1.26}
\end{equation}
so employing $e^{\lambda a^{\dagger}a}=\colon\exp[(e^{\lambda}-1)a^{\dagger
}a]\colon$ and Eq.(\ref{1.10}) ($e^{f}\rightarrow1+\tau$) we see%
\begin{align}
\sum_{m=0}^{\infty}\frac{\tau^{m}}{m!}C_{s,m} & =\left\langle 0(\beta
)\right\vert S_{1}^{\dagger}e^{a^{\dag}a\ln(1+\tau)}S_{1}\left\vert
0(\beta)\right\rangle  \notag \\
& =\left[ C\tau^{2}-2E\tau+1\right] ^{-1/2},  \label{1.27}
\end{align}
where%
\begin{align}
E & =\cosh2r\cosh^{2}\theta-\cosh^{2}r  \notag \\
& =\frac{1}{2}\left[ (2n_{c}+1)\cosh2r-1\right] .  \label{1.28}
\end{align}
Comparing Eq.(\ref{1.27}) with Eq.(\ref{1.17}) yields
\begin{equation}
C_{s,m}=m!C^{m/2}P_{m}\left( E/\sqrt{C}\right) ,  \label{1.29}
\end{equation}
which is the normalization factor of PSSTS. When $r=0$ corresponding to
photon-subtracted thermal state, Eq.(\ref{1.29}) just reduces to $%
C_{s,m}=m!\sinh^{2m}\theta$ \cite{27}.

Using the same procession as obtaining Eq.(\ref{1.24}), the PND of PSSTS is
given by%
\begin{align}
\mathcal{P}_{s}(n) & =C_{s,m}^{-1}\left\langle 0(\beta)\right\vert
S_{1}^{\dagger}a^{\dag m}\left\vert n\right\rangle \left\langle n\right\vert
a^{m}S_{1}\left\vert 0(\beta)\right\rangle  \notag \\
& =\frac{1}{n!}C_{s,m}^{-1}\left\langle 0(\beta)\right\vert S_{1}^{\dagger
}\colon a^{\dag m+n}a^{m+n}e^{-a^{\dag}a}\colon S_{1}\left\vert 0(\beta
)\right\rangle ,  \label{1.30}
\end{align}
so $\left( k=m+n\right) $%
\begin{align}
\sum_{k=0}^{\infty}\frac{\tau^{k}}{k!}n!C_{s,m}\mathcal{P}_{s}(n) &
=\left\langle 0(\beta)\right\vert
S_{1}^{\dagger}e^{a^{\dag}a\ln\tau}S_{1}\left\vert 0(\beta)\right\rangle
\notag \\
& =\text{R.H.S of (\ref{1.21}),}  \label{1.31}
\end{align}
whcih leads to the PND of PSSTS
\begin{equation}
\mathcal{P}_{s}(n)=\frac{\left( m+n\right) !}{n!C_{s,m}\sqrt{A}}\left(
C/A\right) ^{m+n/2}P_{m+n}\left( B/\sqrt{AC}\right) ,  \label{1.32}
\end{equation}
a Legendre polynomial, which is same as the result of Ref.\cite{28}.

\section{Conclusion}

In this paper, we present a new concise approach for normalizing
m-photon-added (-subtracted) STS and for deriving the PNDs, which improve
the method used in Refs. \cite{26,28}. That is to say, using the thermal
field dynamics theory, we convert the thermal state to a pure\ state in
doubled Fock space in which the calculations of ensemble averages under a
mixed state $\rho $, $\left\langle A\right\rangle =\mathtt{tr}\left( A\rho
\right) /\mathtt{tr}\left( \rho \right) \ $is replaced by an equivalent
expectation values with a pure state $\left\vert 0(\beta )\right\rangle $,
i.e., $\left\langle A\right\rangle =\left\langle 0(\beta )\right\vert
A\left\vert 0(\beta )\right\rangle $. It is shown that the average value of $%
e^{fa^{\dag }a}$ under STS is just the generating function of Legendre
polynomials, a remarkable result. Based on this point, the normalization and
PNDs of m-photon added (or subtracted) STS are easily obtained as the
Legendre polynomials. The generating function of the Legendre polynomials
and the average value of $e^{fa^{\dag }a}$ under STS are used in the whole
calculation.

\bigskip

\textbf{ACKNOWLEDGEMENTS:} Work supported by the National Natural Science
Foundation of China (Grant Nos. 11047133, 60978009), the Major Research Plan
of the National Natural Science Foundation of China (Grant No. 91121023),
and the \textquotedblleft 973\textquotedblright\ Project (Grant No.
2011CBA00200), as well as the Natural Science Foundation of Jiangxi Province
of China (No. 2010GQW0027).


\bigskip

\begin{thebibliography}{99}
\bibitem{1} D. Bouwmeester, A. Ekert and A. Zeilinger, The Physics of
Quantum Information (Springer-Verlag, 2000).

\bibitem{2} G. S. Agarwal and K. Tara, Phys. Rev. A 43, 492 (1991); Phys.
Rev. A 46, 485 (1992).

\bibitem{2a} Z. X. Zhang and H. Y. Fan, Phys. Lett. A 165, 14 (1992).

\bibitem{3} A. Zavatta, S. Viciani, and M. Bellini, Science, \textbf{306},
660 (2004).

\bibitem{4} A. Zavatta, V. Parigi, and M. Bellini, Phys. Rev. A \textbf{75},
052106 (2007).

\bibitem{5} A. Zavatta, S. Viciani, and M. Bellini, Phys. Rev. A \textbf{72}%
, 023820 (2005).

\bibitem{6} V. Parigi, A. Zavatta, M. S. Kim, and M. Bellini, Science,
\textbf{317}, 1890 (2007).

\bibitem{7} A. Ourjoumtsev, R. Tualle-Brouri, J. Laurat, Ph. Grangier,
Science \textbf{312}, 83 (2006).

\bibitem{8} A. Ourjoumtsev, A. Dantan, R. Tualle-Brouri, and Ph.Grangier,
Phys. Rev. Lett. 98, 030502 (2007).

\bibitem{9} P. Marek, H. Jeong and M. S. Kim, Phys. Rev. A \textbf{78},
063811 (2008).

\bibitem{10} S. L. Zhang and P. van Loock, Phys. Rev. A \textbf{82}, 06216
(2010).

\bibitem{11} Li-yun Hu and Hong-yi Fan, J. Opt. Soc. Am. B, \textbf{25},
1955 (2008).

\bibitem{11a} Li-yun Hu and Hong-yi Fan, Chin. Phys. B 18, 4657 (2009).

\bibitem{12} T. Opatrn\'{y}, G. Kurizki and D-G. Welsch,\ Phys. Rev. A
\textbf{61,} 032302 (2000).

\bibitem{13} S. Olivares and M. G. A. Paris, J. Opt. B: Quantum
Semiclassical Opt. 7, S392 (2005).

\bibitem{14} A. Kitagawa,M. Takeoka,M. Sasaki, and A. Chefles, Phys. Rev. A
73, 042310 (2006).

\bibitem{15} Y. Yang and F. L. Li, J. Opt. Soc. Am. B, \textbf{26}, 830
(2009).

\bibitem{16} A. Biswas and G. S. Agarwal,\ Phys. Rev. A \textbf{75}, 032104
(2007).

\bibitem{17} S. Y. Lee and H. Nha, Phys. Rev. A 82, 053812 (2010).

\bibitem{18} H. Y. Fan and N. Q. Jiang, Chin. Phys. Lett. \textbf{27},
044206 (2010).

\bibitem{19} Y. Takahashi and H. Umezawa, Collective Phenomena \textbf{2},
55 (1975).

\bibitem{20} Memorial Issue for H. Umezawa, Int. J. Mod. Phys. B \textbf{10}%
, 1695 (1996) memorial issue and references therein.

\bibitem{21} H. Umezawa, Advanced Field Theory-----Micro; Macro; and Thermal
Physics (AIP, 1993).

\bibitem{22} Li-yun Hu and Hong-yi Fan, Chin. Phys. Lett. \textbf{26},
090307 (2009).

\bibitem{23} R. R. Puri, Mathematical Methods of Quantum Optics
(Springer-Verlag, Berlin/Heidelberg/New York, 2001), p. 269 (A.29).

\bibitem{25} George B. Arfken, Hans J. Weber, Mathematical Methods for
Physicists, Elsevier Academic Press, p. 743, (2005).

\bibitem{24} Hongyi Fan, Representation and Transformation Theory in Quantum
Mechanics-----Progress of Dirac's symbolic method (Shanghai Scientific and
technical press, Shanghai, 1997).

\bibitem{26} Li-yun Hu and Z. M. Zhang, [quant-ph] arXiv: 1110.6587.

\bibitem{27} Li-yun Hu and Hong-yi Fan, Mod. Phys. Lett. A \textbf{24}, 2263
(2009).

\bibitem{28} Li-yun Hu, Xue-xiang Xu, Zi-sheng Wang, and Xue-fen Xu, Phys.
Rev. A \textbf{82}, 043842 (2010).
\end{thebibliography}
\end{document}